DOI: 10.1002/marc.201500480Communication

# Porous membranes built up from hydrophilic poly(ionic liquid)s

Karoline Täuber[a], Annett Zimathies[b], Jiayin Yuan[a]*

K. Täuber, Dr. J. Yuan
Max Planck Institute of Colloids and Interfaces, Am Mühlenberg 1, 14476 Potsdam, Germany.
E-mail: Jiayin.yuan@mpikg.mpg.de
A. Zimathies
Federal Institute for Materials Research and Testing, Richard-Willstätter Str. 11, 12489 Berlin, Germany.**Abstract**

Porous polymer membranes *via* electrostatic complexation triggered by neutralization are fabricated for the first time from a water-soluble poly(ionic liquid) (PIL). The porous structure is formed as a consequence of simultaneous phase separation of the PIL and ionic complexation, which occurred in a basic solution of a non-solvent for the PIL. These membranes have a stimuli-responsive porosity, with open and closed pores in isopropanol and in water, respectively. This property is quantitatively demonstrated in filtration experiments, where water is passing much slower through the membranes than isopropanol.TOC

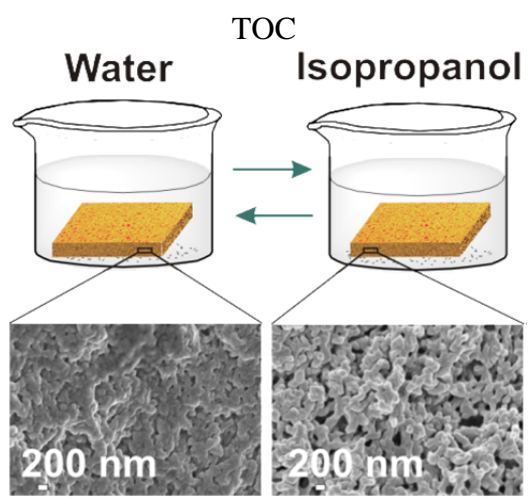

- 1 -

# 1. Introduction

Polyelectrolyte complex (PEC) membranes, obtained *via* ionic crosslinking of polyelectrolytes, are well-known and intensively studied materials that are widely used in industry and research laboratories.[1] Their applications include for example pervaporation,[2] filtration[3] and ion exchange membranes,[4] making them indispensable in the petroleum industry, in potable water purification and in the field of renewable energies, amongst many others.[1g, 2a] From a structural standpoint, porosity acts as a key factor in some of these applications that determines the separation performance and the kinetics of the mass transport across the membranes. In literature PEC membranes bearing micro- or macropores are frequently reported, and nanoporous PEC structures, though synthetically more challenging, are also entering the focus of the porous membrane field.[5] Equally important, membrane wettability in most applications dictates the successful separation and efficiency of the processes.[6] Usually polyelectrolytes incorporated into these membranes are intrinsically of hydrophilic nature thus fabricated in an aqueous environment, a common property of conventional polyelectrolytes. A classic preparation technique of PEC membranes is the robust layer-by-layer (LbL) procedure, where a substrate is dipped subsequently and repeatedly into solutions of oppositely charged polyelectrolytes.[1b-d], [7] This bottom-up method yields membranes of highly controllable thickness, hetero-structure and final charge. Nevertheless, porosity, especially on a nanoscale, may only be induced into these LbL membranes by an extra step, *e.g.* either templating or post-modification methods, leading to the possible use of hazardous etching compounds in the former or partial loss of product in the latter method, not to mention the time consumed.

Our group has previously described a method of how to fabricate (nano)porous membranes through ionic complexation of a hydrophobic cationic poly(ionic liquid) (PIL) with either a polyacid or low molecular weight organic multivalent acids.[8] PILs appear recently as popular candidates for membrane fabrication due to their charged nature and interaction with specific compounds.[9] The uniqueness of this method lies in a simultaneous network- and pore-formation process, in which aqueous ammonia solution diffused through the PIL/acid blend film sticking to a glass substrate from one side to the other. Various evidences indicate that the pores were built up due to a phase-separation mechanism of the ionic, hydrophobic PIL in an aqueous environment; the nanostructure that formed during the phase separation is *in-situ* locked by ionic complexation between carboxylates and the cationic PIL. The success of this technique relies on, to our opinion, the necessary employment of PILs of certain hydrophobicity that will undergo structure rearrangement in water.[8, 9f, g] This assumption is supported by the fact that our efforts to apply PILs bearing halide anion, thus of high hydrophilicity, for this membrane fabrication technique failed to produce porous membranes.

In this contribution, we introduce a modified approach of our previous membrane fabrication procedure to enable the processing of water-soluble PILs into hydrophilic porous membranes. *Via* using an organic solvent-dominant solution instead of water in the membrane fabrication process to form porous networks through phase separation, we have now been able to fabricate membranes from hydrophilic PILs, which have different properties from our previously described porous hydrophobic PIL-membranes. These membranes hold a stimuli-responsive porosity, *i.e.* the pore structure changes from open to close in response to solvents of different polarity due to swelling.

## 2. Experimental Section

Experimental details including the synthesis of the polymers and the fabrication and analysis of the membranes are included in the supporting information.

## 3. Results and Discussion

### 3.1. Preparation of the membrane

The porous membranes were initially prepared from a water-soluble vinylimidazolium based PIL bearing Br⁻ as counteranion, poly(3-cyanomethyl-1-vinylimidazolium bromide) (PCMVIm-Br). It was ionically crosslinked by carboxylates of benzoic acid derivatives carrying multiple acid groups. The synthetic route is displayed in **Figure 1**. Experimentally, the PIL and the multivalent acid were first dissolved homogenously in dimethyl sulfoxide (DMSO), a polar aprotic solvent that prevents the charge state of the acid compound, thus the ionic complexation of the PIL and the acid upon direct mixing. The mixture solution was then cast into a thin film on a glass slide and the solvent DMSO was evaporated overnight at 80 °C. The dry polymer/acid blend film sticking to the glass slide was subsequently immersed into a complexation solution, *i.e.* a mixture liquid containing 0.7 wt% of base (KOH or $NH_3$) and 8 vol% of water in isopropanol (**Figure 1**). After 2 h, the formed membrane was detached easily from the glass slide on account of its porous membrane structure that reduced its interaction with the glass substrate. In order to show that this membrane formation method is generally applicable to other hydrophilic PILs, we have also prepared porous membranes of similar structures from a pyridinium based PIL bearing Br⁻ as counter anion (**Figure S4**).

### 3.2. Crosslinking and pore formation

The concurrent penetration of base and solvent molecules into the polymer/acid blend film sticking firmly to a glass substrate triggers a deprotonation-induced complexation between the carboxylate and the cationic polyvinylimidazolium; at the same time the PIL is structurally rearranging in the complexation solution dominated by isopropanol, a non-solvent for the polymer. The simultaneous occurrence of phase separation and ionic complexation results in the formation of a porous network locked by ionic crosslinking, that is, the structure of the phase separating PIL is immediately in-situ frozen by the ionic crosslinking with the carboxylates.

The porous structure of PCMVIm-Br membranes appears very much different from what we usually found in previously reported membranes built up from a hydrophobic PIL bearing TFSI (TFSI: bis(trifluoromethylsulfonyl)imide)) as counteranion.[8] In comparison to the previously reported membranes carrying a porous structure inside a polymeric matrix, we now have a rather open porous structure where the polymer network is surrounded by free space (**Figure S5**). This new porous structure is expected to be related to the difference in the phase separation process of the two types of membranes during formation. Lacking of hydrophobic interactions of the PCMVIm-Br polymer chains during the current phase separation step, which were present in the membrane fabrication using hydrophobic PILs, the diffusion of the complexation solution into the polymer/acid blend film is accelerated, forming pores that are significantly increased in size.

Investigation of the membrane's porous structure by mercury intrusion revealed a broad monomodal pore size distribution. Considering the mercury intrusion at different pressures, two mercury uptakes, one at medium pressures and one at high pressures, were observed. The mercury uptake at pressures above 50 MPa indicates the possible presence of meso- or micropores. Nitrogen sorption measurements revealed however that pores with size below 50 nm are absent or negligible. These two contradictive observations may stem from elastic deformations of the porous polymer network at high pressures during the mercury intrusion. Therefore, the pore size distribution and porosity of the membrane was analyzed by mercury intrusion experiment up to 50 MPa. The membranes had a broad pore size distribution from 100 nm to 2 µm, a modal (most frequent) pore diameter of 700 nm and a porosity of 38% (**Figure 2**).

### 3.3. Influence of water in the complexation solution

We have investigated the influence of the addition of water to the complexation solution using NH$_3$ as base to study the phase separation and membrane formation mechanism. Addition of small amounts of water to the complexation solution is necessary to obtain isolatable membranes from the substrate. Meanwhile, water as a good solvent for PCMVIm-Br, hinders the phase separation process and therefore inhibits the pore formation. In order to find the optimal amount of water, membranes were prepared in solutions of isopropanol/$NH_3$ with various concentrations of water (**Table 1, Figure S6**). We found that the overall degree of electrostatic complexation (DEC, the percentage of cross-linked repeating units of the PIL, which was studied by the elemental analysis of $Br^-$, since $Br^-$ is released into solution upon ionic complexation) is increasing from 20 % without addition of water to 60 % by addition of 23 vol% of water. The high DEC is related to an improved solubility of the complexation product $NH_4Br$ in isopropanol in the presence of more water, making the crosslinking entropically more favorable. This high DEC is accompanied by a change in the physical state of the membrane. Without addition of water, the pore structure is more finely constructed (**Figure S6A**); but instead of large-sized membranes, only small broken pieces were isolated on the glass plate. At too high water concentrations, mechanically stable and free-standing membranes formed, but with closed larger pores with size above 5 µm. This phenomenon is related to an incompletely developed phase separation of the hydrophilic PIL in the presence of more water. As a good compromise between the membrane integrity and its porous structure, the water content in the complexation solution was set to 8 vol % for the remaining experiments.

### 3.4. Influence of the base on the network formation

Studying the influence of the base molecules on the properties of the resulting membrane, two different candidates, KOH and $NH_3$, were chosen. Structurally, membranes prepared from KOH and $NH_3$ appear similar in scanning electron microscopy (SEM) characterization. However, a clear difference is found in the DEC of the entire membranes. We surprisingly found that all membranes prepared from KOH presented a much higher DEC (above 82%) than these from $NH_3$ (around 20%). This apparent higher DEC in the case of KOH is caused by an intensive anion exchange process ($c_{(KOH)} \sim 1.5$ M) from $Br^-$ to $OH^-$ simultaneous to the ionic complexation during the base molecule diffusion, eliminating $Br^-$ from the membrane. From this standpoint, the DEC value obtained from membranes prepared in $NH_3$ solution is more reliable ($c_{(OH^-)} \sim 1 \times 10^{-3}$ M).

### 3.5. Stimuli-responsive porosity

The difference in the membrane fabrication method compared to the previously described PIL-membranes results in new properties. These hydrophilic PCMVIm-Br membranes swell in water on account of the hydrophilic nature of the polymer. Due to their relatively lower crosslinking degree (DEC ~ 20%) than previously reported membranes (DEC ~ 50%),[8b] PCMVIm-Br-membranes are obviously responding to solvent change, for instance from water to isopropanol by reversible pore closing and opening (**Figure 3**). When soaked in water, the PIL chains, which did not undergo ionic complexation (ca. 80% of overall PIL chains according to the DEC value), swell and the membrane pores will be blocked, whereas in isopropanol, a non-solvent for the PCMVIm-Br, the polymer tries to minimize its interactions with the solvent, and therefore takes a de-swelling conformation, *i.e.* the pores open again.

The reversible opening and closing of the membrane pores has been demonstrated in filtration experiments using isopropanol and water as eluent. The solvent flow through the membrane is 5.41 mL $cm^{-2}s^{-1}$ for isopropanol, 2.5 times of the flow observed for water, which is 2.14 mL $cm^{-2}s^{-1}$. Showing the repeatability of the responsive behavior of the porous structure, we have performed subsequent water and isopropanol flow experiments without significant changing of the flow of the solvents after 10 cycles (**Figure 3**).

### 4. Conclusions

In summary, we have introduced a new fabrication method of how to access porous membranes from hydrophilic poly(ionic liquid)s, combining simultaneous pore formation and ionic complexation. It is demonstrated that these membranes show reversible pore opening and closing behavior which has been used to modulate the solvent flow in a filtration test. This new function makes these membranes potentially applicable as stimuli-responsive filtration systems, smart sensors or controlled loading and release systems.

## Supporting Information

Supporting Information is available from the Wiley Online Library or from the author

## Appendix/Nomenclature/Abbreviations

Acknowledgements: This research was supported by Max Planck Society in Germany and the ERC (European Research Council) Starting Grant 2014 with project number 639720 – NAPOLI.

Received: Month XX, XXXX; Revised: Month XX, XXXX; Published online:
DOI: 10.1002/marc.((insert number))

**Keywords**: poly(ionic liquid), porous membranes, crosslinking, phase separation, responsive materials

[1]     (a) Chen, H.; Palmese, G. R.; Elabd, Y. A., *Chem. Mater.* **2006,** *18* (20), 4875-4881; (b) Decher, G.; Hong, J. D.; Schmitt, J., *Thin Solid Films* **1992,** *210–211, Part 2* (0), 831-835; (c) Dodoo, S.; Steitz, R.; Laschewsky, A.; von Klitzing, R., *Phys. Chem. Chem. Phys.* **2011,** *13* (21), 10318-10325; (d) Laschewsky, A.; Mayer, B.; Wischerhoff, E.; Arys, X.; Jonas, A.,

*Ber. Bunsen-Ges. Phys. Chem.* **1996,** *100* (6), 1033-1038; (e) Thünemann, A.; Müller, M.; Dautzenberg, H.; Joanny, J.-F.; Löwen, H., Polyelectrolyte Complexes. In *Polyelectrolytes with Defined Molecular Architecture II*, Schmidt, M., Ed. Springer Berlin Heidelberg: 2004; Vol. 166, pp 113-171; (f) v. Klitzing, R.; Tieke, B., Polyelectrolyte Membranes. In *Polyelectrolytes with Defined Molecular Architecture I*, Schmidt, M., Ed. Springer Berlin Heidelberg: 2004; Vol. 165, pp 177-210; (g) Zhao, Q.; An, Q.; Ji, Y. L.; Qian, J. W.; C:J., G., *J. Membr. Sci.* **2011,** *379*, 19-45; (h) Zhumadilova, G. T.; Gazizov, A. D.; Bimendina, L. A.; Kudaibergenov, S. E., *Polymer* **2001,** *42* (7), 2985-2989.

[2] (a) Baker, R. W., *J. Membr. Sci.* **2010,** *362* (1–2), 134-136; (b) Chen, Y.; Xiangli, F.; Jin, W.; Xu, N., *J. Membr. Sci.* **2007,** *302* (1–2), 78-86; (c) Krasemann, L.; Tieke, B., *J. Membr. Sci.* **1998,** *150* (1), 23-30.

[3] (a) Lutkenhaus, J. L.; McEnnis, K.; Hammond, P. T., *Macromolecules* **2008,** *41* (16), 6047-6054; (b) Sanyal, O.; Sommerfeld, A. N.; Lee, I., *Sep. Purif. Technol.* (0); (c) Tyagi, P.; Deratani, A.; Bouyer, D.; Cot, D.; Gence, V.; Barboiu, M.; Phan, T. N. T.; Bertin, D.; Gigmes, D.; Quemener, D., *Angew. Chem., Int. Ed.* **2012,** *51* (29), 7166-7170.

[4] (a) Kraytsberg, A.; Ein-Eli, Y., *Energy Fuels* **2014,** *28* (12), 7303-7330; (b) Merle, G.; Wessling, M.; Nijmeijer, K., *J. Membr. Sci.* **2011,** *377* (1–2), 1-35; (c) Steele, B. C. H.; Heinzel, A., *Nature* **2001,** *414* (6861), 345-352.

[5] (a) Dotzauer, D. M.; Dai, J.; Sun, L.; Bruening, M. L., *Nano Lett.* **2006,** *6* (10), 2268-2272; (b) Guo, W.; Tian, Y.; Jiang, L., *Acc. Chem. Res.* **2013**; (c) Li, Q.; Quinn, J. F.; Caruso, F., *Adv. Mater. (Weinheim, Ger.)* **2005,** *17* (17), 2058-2062; (d) Mendelsohn, J. D.; Barrett, C. J.; Chan, V. V.; Pal, A. J.; Mayes, A. M.; Rubner, M. F., *Langmuir* **2000,** *16* (11), 5017-5023; (e) Reisch, A.; Tirado, P.; Roger, E.; Boulmedais, F.; Collin, D.; Voegel, J.-C.; Frisch, B.; Schaaf, P.; Schlenoff, J. B., *Adv. Funct. Mater.* **2013,** *23* (6), 673-682; (f) Tokarev, I.; Minko, S., *Adv. Mater. (Weinheim, Ger.)* **2010,** *22* (31), 3446-3462; (g) van Rijn, P.; Tutus, M.;

Kathrein, C.; Zhu, L.; Wessling, M.; Schwaneberg, U.; Boker, A., *Chem. Soc. Rev.* **2013,** *42* (16), 6578-6592.

[6]     Sirkar, K. K., *Ind. Eng. Chem. Res.* **2008,** *47* (15), 5250-5266.

[7]     (a) Bertrand, P.; Jonas, A.; Laschewsky, A.; Legras, R., *Macromol. Rapid Commun.* **2000,** *21* (7), 319-348; (b) Decher, G.; Hong, J.-D., *Makromolekulare Chemie. Macromolecular Symposia* **1991,** *46* (1), 321-327; (c) Koetse, M.; Laschewsky, A.; Jonas, A. M.; Wagenknecht, W., *Langmuir* **2002,** *18* (5), 1655-1660; (d) Lvov, Y.; Decher, G.; Moehwald, H., *Langmuir* **1993,** *9* (2), 481-486; (e) Mallwitz, F.; Laschewsky, A., *Adv. Mater. (Weinheim, Ger.)* **2005,** *17* (10), 1296-1299.

[8]     (a) Täuber, K.; Zhao, Q.; Antonietti, M.; Yuan, J., *ACS Macro Letters* **2015,** *4*, 39-42; (b) Zhao, Q.; Dunlop, J. W. C.; Qiu, X.; Huang, F.; Zhang, Z.; Heyda, J.; Dzubiella, J.; Antonietti, M.; Yuan, J., *Nat Commun* **2014,** *5*; (c) Zhao, Q.; Heyda, J.; Dzubiella, J.; Täuber, K.; Dunlop, J. W. C.; Yuan, J., *Adv. Mater. (Weinheim, Ger.)* **2015**, n/a-n/a; (d) Zhao, Q.; Yin, M.; Zhang, A. P.; Prescher, S.; Antonietti, M.; Yuan, J., *J. Am. Chem. Soc.* **2013,** *135* (15), 5549-5552; (e) Täuber, K.; Lepenies, B.; Yuan, J., *Polymer Chemistry* **2015,** *6* (27), 4855-4858.

[9]     (a) Carlisle, T. K.; Wiesenauer, E. F.; Nicodemus, G. D.; Gin, D. L.; Noble, R. D., *Ind. Eng. Chem. Res.* **2013,** *52* (3), 1023-1032; (b) Nguyen, P. T.; Voss, B. A.; Wiesenauer, E. F.; Gin, D. L.; Noble, R. D., *Ind. Eng. Chem. Res.* **2013,** *52* (26), 8812-8821; (c) Qiu, B.; Lin, B.; Qiu, L.; Yan, F., *J. Mater. Chem.* **2012,** *22* (3), 1040-1045; (d) Si, Z.; Sun, Z.; Gu, F.; Qiu, L.; Yan, F., *Journal of Materials Chemistry A* **2014,** *2* (12), 4413-4421; (e) Tome, L. C.; Mecerreyes, D.; Freire, C. S. R.; Rebelo, L. P. N.; Marrucho, I. M., *Journal of Materials Chemistry A* **2014,** *2* (16), 5631-5639; (f) Yang, Y.; Zhang, Q.; Li, S.; Zhang, S., *RSC Advances* **2015,** *5* (5), 3567-3573; (g) Zhang, K.; Feng, X.; Sui, X.; Hempenius, M. A.; Vancso, G. J., *Angew. Chem., Int. Ed.* **2014,** *53* (50), 13789-13793, (h) D. Mecerreyes, *Prog.*

*Polym. Sci.* **2011**, *36*, 1629, (i) J. Lu, F. Yan, J. Texter, *Prog. Polym. Sci.* **2009**, *34*, 431, (j) J. Yuan, D. Mecerreyes, M. Antonietti, *Prog. Polym. Sci.* **2013**, *38,* 1009.

Figures

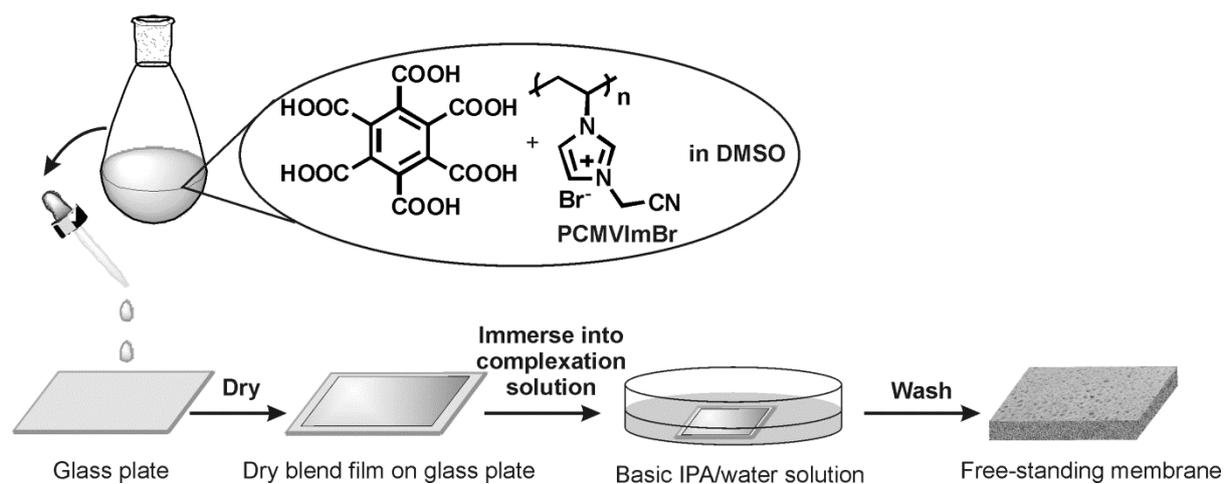

*Figure 1.* General fabrication procedure of porous PIL membranes on the example of PCMVIm-Br and mellitic acid.

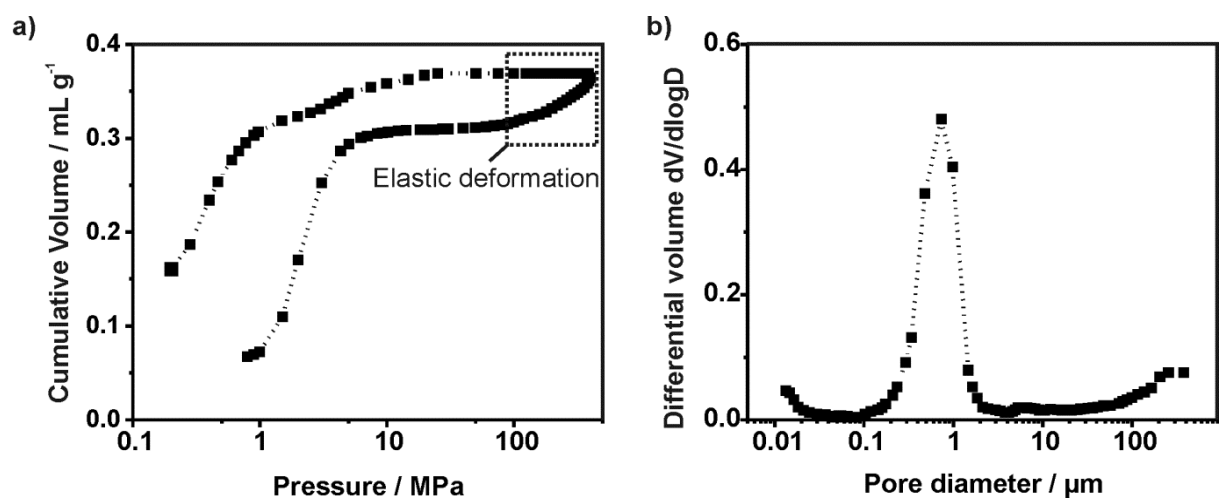

*Figure 2.* Results of porosity studies by mercury intrusion. a) Cumulative volume of mercury versus pressure. b) Pore size distribution curve of the porous membrane prepared in a complexation solution with 8 vol% of water.

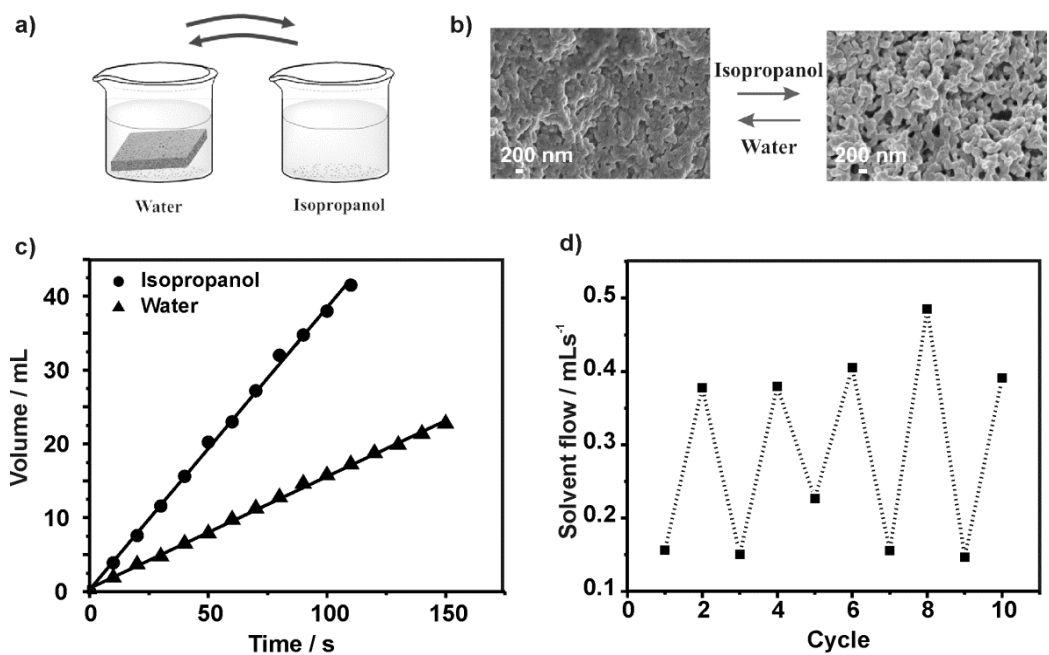

*Figure 3.* a) Schematic representation of the solvent experiment by reversibly immersing the membrane in water and isoproposal. b) SEM pictures of the membrane after water soaking (left) and isopropanol soaking (right). c) Recording of the solvent flow through the membranes as a function of time for water and isopropanol. d) Subsequent recording of water and isopropanol flow through the membrane.

*Table 1.* Comparison of the degree of electrostatic complexation (DEC) of the membranes with different amounts of water in the isopropanol soaking solution.

| Amount of water (vol%) | 0 | 8 | 23 |
|---|---|---|---|
| DEC (%) for NH3 soaking | 20 | 16 | 60 |

**Porous polymer membranes were prepared from hydrophilic poly(ionic liquid)s by concurrent phase separation and ionic crosslinking.** We investigated the pore formation mechanism and showed that these membranes have a stimuli-responsive porosity. The pores are open in isopropanol and closed in water, which was demonstrated in filtration experiments.

**K. Täuber, A. Zimathies, J. Yuan***

**Porous membranes from hydrophilic poly(ionic liquid)s**

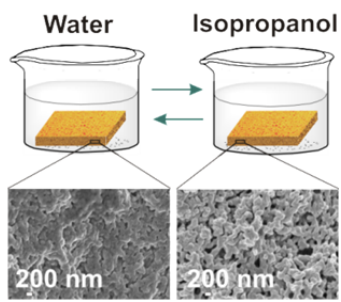

# Supporting Information

for *Macromol. Rapid Commun.*, DOI: 10.1002/marc.201500480

**Porous membranes built up from hydrophilic poly(ionic liquid)s**

K. Täuber, A. Zimathies, J. Yuan*

## 1. Experimental procedures

### 1.1. Materials

Mellitc acid (99%), 1,2,4,5-tetracarboxylic benzoic acid (96%), 1-vinylimidazole (99%) and poly(4-vinyl pyridine) ($M_w$ = 60 kDa) were purchased from Sigma Aldrich, 1,2,3,4,5-Pentacarboxylic benzoic acid was purchased from TCI; 1,2,3-Tricarboxyl benzoic acid (99%) was purchased from Alfa Aesar; Isopthtalic acid (99%) was purchased from Acros. All chemicals were used without further purification. The solvents were of analytical grade.

### 1.2. Synthesis of PCMVIm-Br

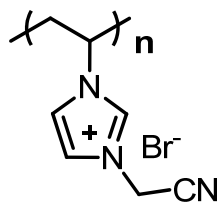

**PCMVIm-Br**

**PCMVIm-Br** was synthesized as previously reported.[1] Its chemical structure was confirmed by proton nuclear magnetic resonance ($^1$H-NMR) and is shown in Figure S1. The molecular weight was determined to be 54 kDa and the PDI to be 3.38 by GPC measurements (eluent: water + acetate buffer, 20% MeOH) using PEG as standard.

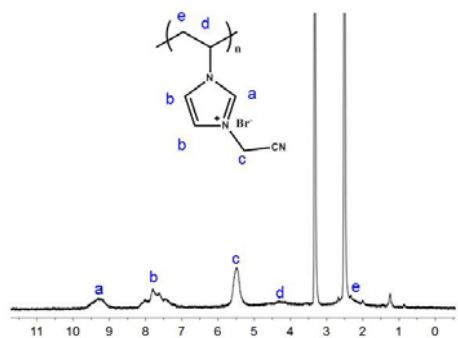

**Figure S1**: Proton nuclear magnetic resonance (400 Hz) spectrum of **PCMVIm-Br** using DMSO-$d_6$ as NMR solvent. The chemical shifts at 2.5 and 3.3 ppm are from residual DMSO and H$_2$O.

### 1.3. Synthesis of PCMPy-Br

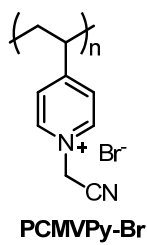

PCMVPy-Br

PCMVPy-Br was synthesized as described earlier.[2] Its chemical structure was confirmed by $^1$H-NMR as shown in Figure S2. As it was synthesized from commercially available P4VP ($M_w$ = 60 kDa), its $M_w$ was calculated to be 128 kDa.

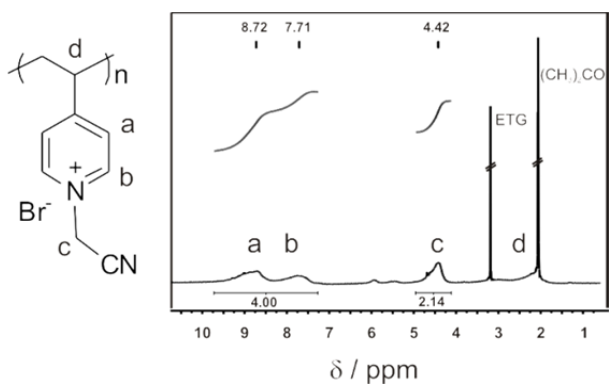

**Figure S2**: Proton nuclear magnetic resonance spectrum of **PCMVPy-Br** using acetone-$d_6$ as NMR solvent. The chemical shifts at 2 and 3.2 ppm are from residual acetone and ethylene glycol (ETG), which is a solvent used for the quaternization reaction.

## 1.4. Membrane fabrication

The membranes were prepared from a homogenous solution of the respective polymer (0.25 mmol) and 1 equivalent of acid in terms of the molar ratio between the imidazolium unit to the carboxyl group. The solution was poured onto a glass plate, and the solvent was evaporated overnight in an oven at 80 °C. The as-obtained dried sticky film was immersed into a solution of 0.7 wt% of KOH in isopropanol containing additional 8 vol% of water. After 2 hours of soaking, the membrane had detached itself from the glass plate and could be isolated as a free-standing membrane and washed with water and isopropanol.

## 1.5. Filtration experiments

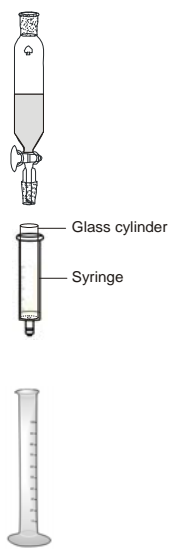

The membrane was attached to a hollow glass cylinder with a hole in the bottom (Ø 1.5 mm). The glass cylinder was then put into an empty syringe and solvents were passing through the device. The glass cylinder was added in order to make sure that the solvent does not flow through the syringe wall without passing by the membrane. In order to keep the weight of the solvent on top of the membrane constant, a separation funnel was placed above the syringe setup and the solvent flow into the syringe was adjusted to keep a constant weight. The volume of solvent passing through the membrane was measured by collecting it in a measuring cylinder.

**Figure S3**: Set up for quantification of solvent flow through the membrane.

## 1.6. Mercury intrusion

The porosity and the pore size distributions of the membrane were measured by a mercury intrusion porosimeter for meso- and macropore analysis with an Autopore III device (Micrometrics, USA) according to DIN 66133 at BAM Federal Institute for Materials Research and Testing, Berlin.

## 1.7. Determination of degree of electrostatic complexation (DEC)

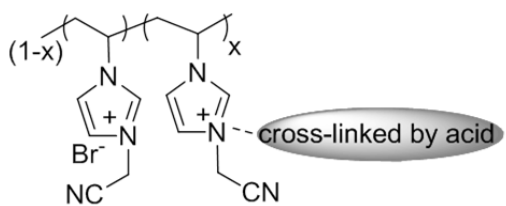

Elemental analysis of bromide was used to determine the DEC. The following formula was applied:

$$DEC = \frac{310x - 80}{120x - 80}$$

Where x equals the amount of bromide in wt% / 100.

## 2. Additional Data

## 2.1. Membrane from pyridinium based PIL

The membranes prepared from PCMVPy-Br were fabricated as described above for PCMVIm-Br membranes with mellitic acid as crosslinker.

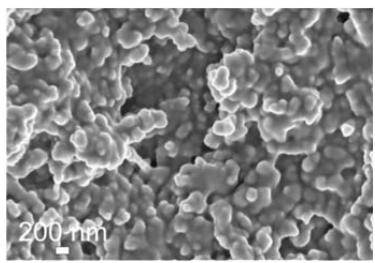

**Figure S4:** Cross-sectional SEM image of a membrane from PCMVPy-Br and mellitic aicd.

## 2.2. Comparison of porous structures between hydrophobic and hydrophilic PCMVIm –PIL membranes

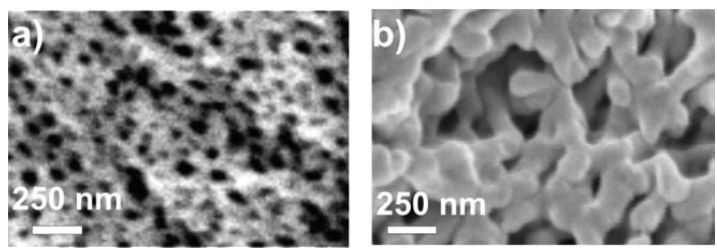

**Figure S5**: Comparison of the pore structure between a) hydrophobic PCMVIm-TFSI[3] and b) hydrophilic PCMVIm-Br membranes.

## 2.3. Influence of amount of water in complexation solution on porous structure

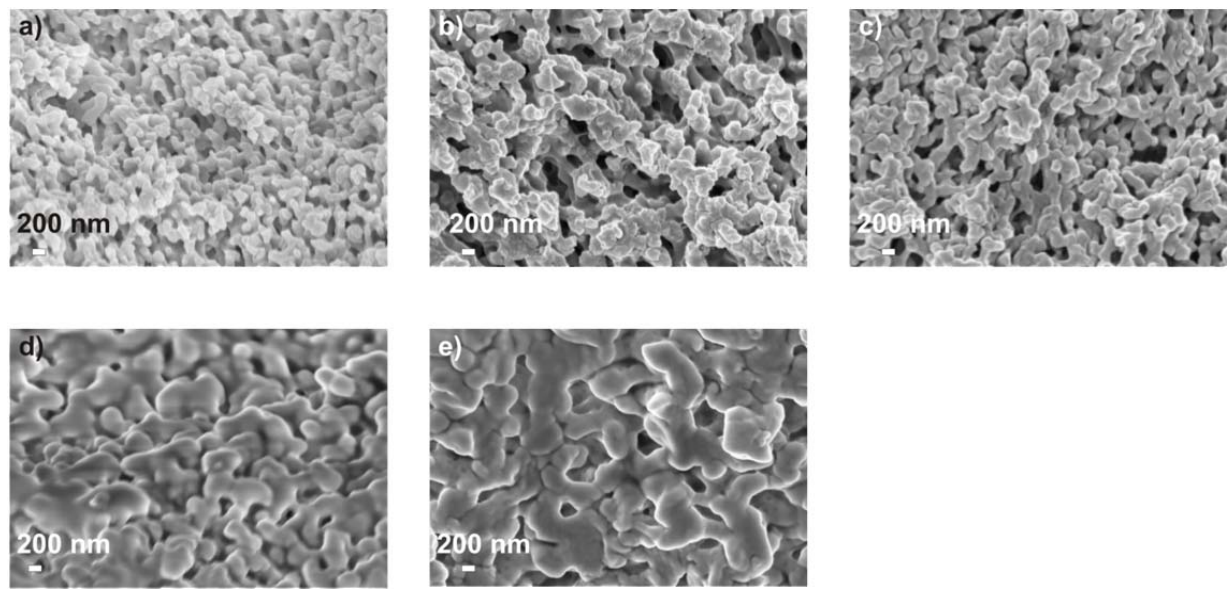

**Figure S6:** SEM images of the cross-sections of the membranes for different amounts of water in the soaking solution: a) 0 vol%. b) 2 vol%. c) 8 vol%. d) 12 vol%. e) 23 vol%.

## 2.4. Influence of acid's multivalency on porous structure

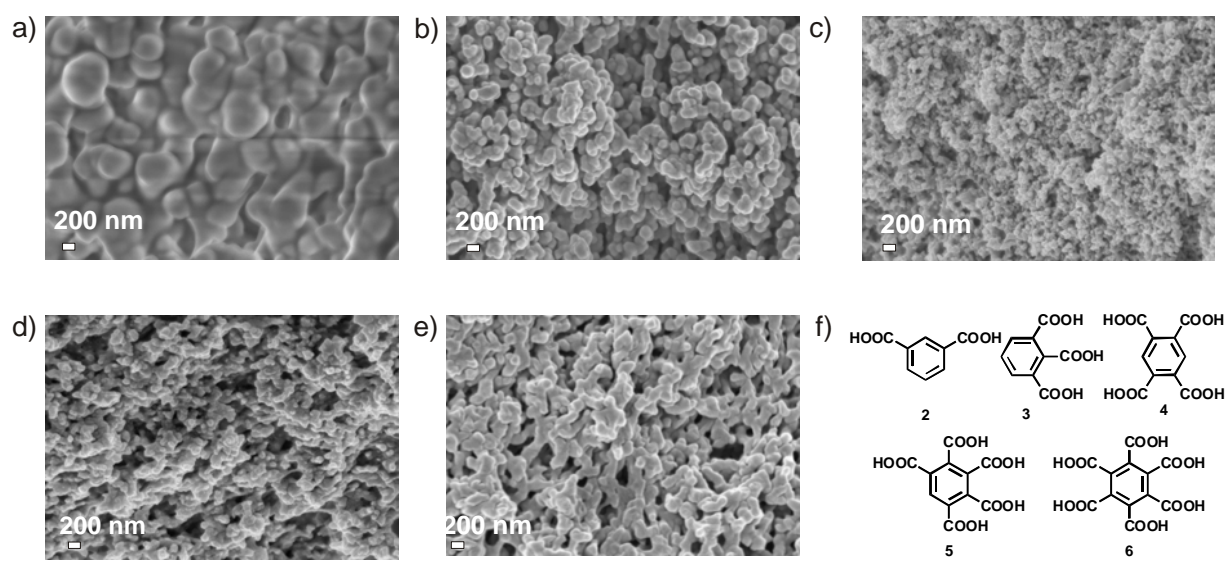

**Figure S7**: SEM Images of membranes made from different acids shown in f). a) acid **2** b) acid **3**. c) acid **4**. d) acid **5**. e) acid **6.** The porous structure is becoming increasingly open as acids of higher multivalency are applied for crosslinking.

[1]     Yuan, J.; Giordano, C.; Antonietti, M., *Chem. Mater.* **2010,** *22* (17), 5003-5012.

[2]     Yuan, J.; Marquez, A. G.; Reinacher, J.; Giordano, C.; Janek, J.; Antonietti, M., *Polym. Chem.* **2011,** *2* (8), 1654-1657.

[3]     Täuber, K.; Zhao, Q.; Antonietti, M.; Yuan, J., *ACS Macro Letters* **2015,** *4*, 39-42.